\documentclass[10pt,a4paper]{article}

\usepackage[latin1]{inputenc}
\usepackage{fourier}
\usepackage{graphicx}
\usepackage{amsmath}
\usepackage{amsfonts}
\usepackage[latin1]{inputenc}
\usepackage[T1]{fontenc}
\usepackage{verbatim}
\usepackage{latexsym}
\usepackage{natbib}

\usepackage{hyperref}
\hypersetup{backref,colorlinks=true,citecolor=blue}

\begin{document}

\title{Coronal Temperature Maps from Solar EUV images: \\ a Blind Source Separation Approach}

\author{T. Dudok de Wit$^{1}$, S. Moussaoui$^{2}$, C. Guennou$^{3}$, F. Auch\`ere$^{3}$, G. Cessateur$^{1,4}$, \\
M. Kretzschmar$^{5}$, L.~A. Vieira$^{1}$, F.~F. Goryaev$^{6}$ \\
\small $^{1}$ LPC2E, 3A avenue de la Recherche Scientifique, 45071 Orl\'eans Cedex 2, France \\ 
\small $^{2}$ IRCCYN, 1 rue de la No\'e, BP 92101, 44321 Nantes Cedex 3, France \\
\small $^{3}$ IAS, B\^atiment 121, 91405 Orsay, France \\
\small $^{4}$ PMOD-WRC, 7260 Davos Dorf, Switzerland \\
\small $^{5}$ SIDC, Royal Observatory of Belgium, Ringlaan 3, 1180 Brussels, Belgium \\
\small $^{6}$ Lebedev Physical Inst., Russian Academy of Sciences, Leninskii pr. 53, Moscow, 119991 Russia}

\date{\normalsize \textit{Article to appear in Solar Physics (2012)}}

\maketitle

\begin{abstract}
Multi-wavelength solar images in the EUV are routinely used for ana\-ly\-sing solar features such as coronal holes, filaments, and flares. However, images taken in different bands often look remarkably similar as each band receives contributions coming from regions with a range of different temperatures. This has motivated the search for empirical techniques that may unmix these contributions and concentrate salient morphological features of the corona in a smaller set of less redundant \textit{source images}. Blind Source Separation (BSS)  precisely does this. Here we show how this novel concept also provides new insight into the physics of the solar corona, using observations made by SDO/AIA. The source images are extracted using a Bayesian positive source separation technique. We show how observations made in six spectral bands, corresponding to optically thin emissions, can be reconstructed by linear combination of three sources. These sources have a narrower temperature response and allow for considerable data reduction since the pertinent information from all six bands can be condensed in only one single composite picture. In addition, they give access to empirical temperature maps of the corona. The limitations of the BSS technique and some applications are briefly discussed.
\end{abstract}

\section{Introduction}

Multiwavelength solar Extreme-UV (EUV) observations are widely used for imaging the complex structure of the solar corona, but are also useful for inferring quantitative information about the thermodynamic state of the solar atmosphere, in particular the density and the temperature. This inference, however, requires radiation transfer models, and is often done at the expense of strong assumptions such as local thermodynamic equilibrium \citep{phillips08}. A conceptually different, and certainly less explored, road involves empirical methods that are generally faster but instead provide more qualitative information. 

Here, we explore such an approach, called Blind Source Separation (BSS), which has recently become a very fertile area of research in various disciplines such as speech processing, biomedical imaging, chemometrics, and remote sensing \citep{comon10,kuruoglu10}. Given a series of linearly mixed signals, BSS provides a framework for recovering the original sources that these observations are made of, using the least prior information. 

The \textit{Atmospheric Imaging Assembly} \citep{lemen12} onboard the recently launched \textit{Solar Dynamics Observatory} (SDO/AIA) routinely observes the solar atmosphere in ten spectral bands, six of which mostly contain optically thin coronal lines. Images taken in these six bands are highly correlated (see Figure 1) and often the information on the underlying physics is found in the subtle difference between various spectral bands \citep{depontieu11}. 

\begin{figure}[!t]
\centerline{\includegraphics[width=1.0\textwidth,clip=]{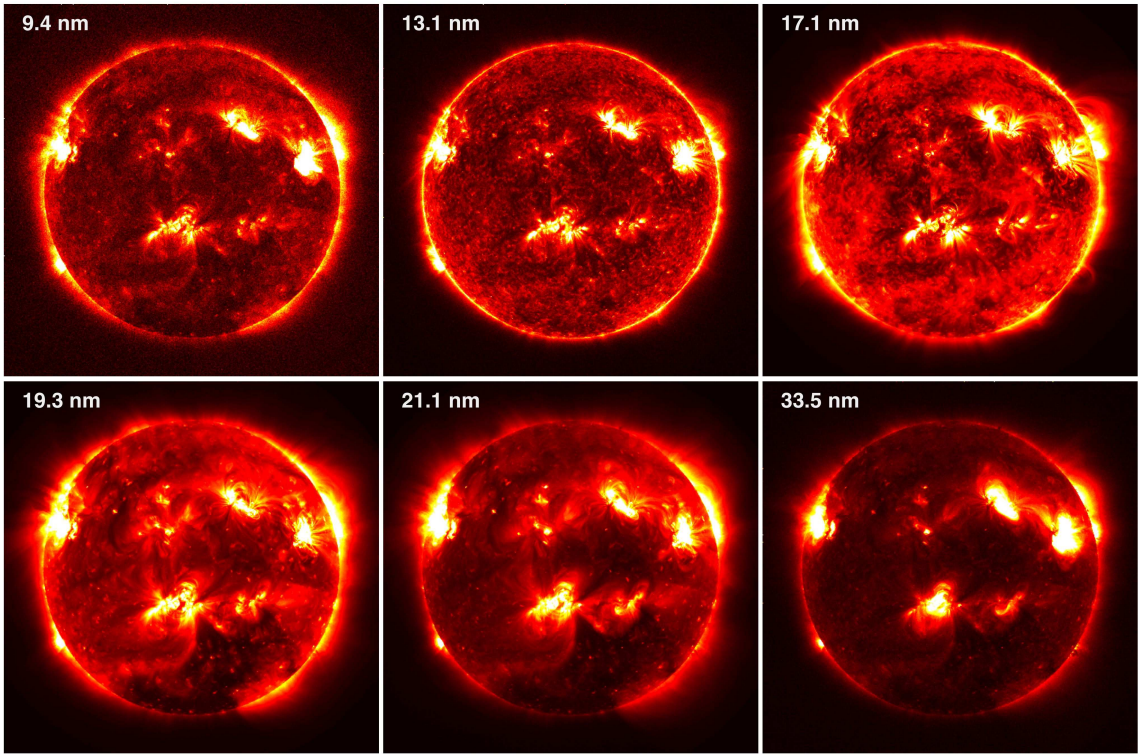}}
\caption{Images taken in six wavelengths by AIA on 12 March 2011 at 23:55 UT. Intensities are shown between the 0.01 and the 0.99 quantiles. In what follows, all intensity images will be corrected for low luminance by applying a gamma correction of $\gamma = 0.7$.}
\label{Figraw}
\end{figure}

\begin{figure}[!b]
\centerline{\includegraphics[width=0.7\textwidth,clip=]{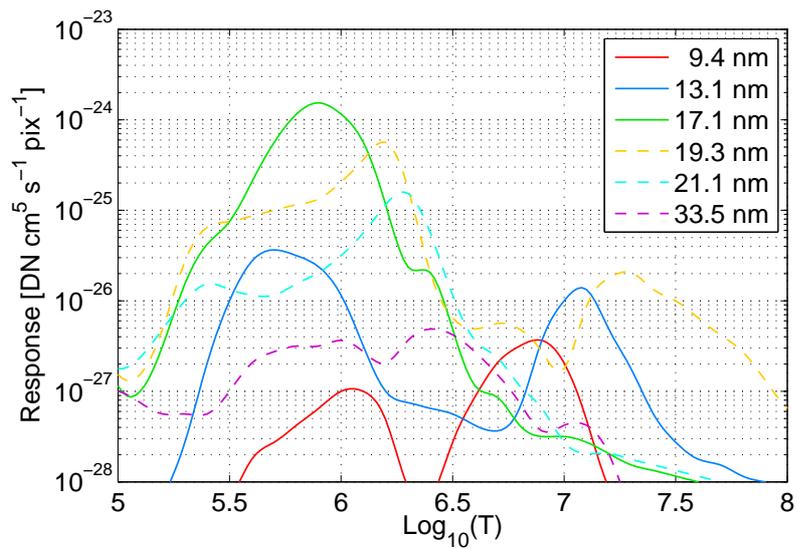}}
\caption{Response of six spectral bands of AIA calculated from the effective area functions and using a \textsf{CHIANTI} model of solar emissivity.}
\label{Figresponse}
\end{figure}

There are two main reasons for this high correlation: \textit{i}) the temperature response associated with each spectral line is generally wide, and sometimes even multimodal; \textit{ii}) because of the finite spectral resolution of the instrument, each spectral band captures a blend of different lines, hence the simultaneous presence of emissions originating both from cold and hot regions.  Images taken in spectral lines with different ionization states also end up being correlated when the temperature responses of these lines overlap. This is illustrated in Figure~\ref{Figresponse} by the instrument response \citep{boerner12}, as calculated using a \textsf{CHIANTI} model of solar emissivity. Note in particular how bands that capture active regions and flares ($\log_{10}(T) > 6.5$) also receive a significant fraction of emissions from lower-temperature plasmas. A spectral unmixing is required to retrieve the individual contributions from these mixtures. 

Many attempts have been made to characterize the solar atmosphere from  a limited set of images that have a broad temperature response and are subject to calibration issues.  The classical approach involves a physical modelling of the Differential Emission Measure (DEM), see for example  \citet{golub04}. There is a growing interest, however, for empirical approaches that are faster and less dependent on the calibration of the images, at the expense of being less accurate. Taking the difference between images in two different wavelengths is a common trick for enhancing emissions that originate either from the hot, or from the cool corona \citep{reale11}. Of course,  better separations would be achieved by combining more than just two images. The apparent lack of methodology, however, for finding such combinations, has prevented this approach from being further developed. Here, we pursue this idea and introduce a rigorous statistical framework for determining appropriate combinations of images.

The BSS model assumes that each image [$I$], expressed by its pixel intensities (labelled as a function of position $x$ and wavelength $\lambda$), can be decomposed into a weighted sum of source images [$S_k(x)$] 
\begin{equation}
I(\lambda,x) = \sum_k V_k(\lambda) S_k(x) + B(\lambda,x) ,
\label{bss1}
\end{equation}
whose weights [$V_k(\lambda)$] are called mixing coefficients; $B(\lambda,x)$ is a noise term that models measurement errors and model uncertainties. In BSS, both $V(\lambda)$ and $S(x)$ must be inferred from  $I(\lambda,x)$ only. This inverse problem is severely ill-posed, so prior knowledge is needed to constrain the solution to be unique. In the following, we require the sources and their mixing coefficients to be positive, and mutually independent in a probabilistic sense. As we shall see, the resulting source images are remarkably close to our physical perception of what the individual solar contributions should be. In particular, they can be used to rapidly infer information on the thermal structuring of the solar corona.

The data and the BSS technique are respectively presented in Sections 2 and 3. The analysis procedure and the interpretation of the results are discussed in Sections 4 to 6. Outlooks and conclusions follow in Sections 7 and 8.


\section{The Data and Physical Assumptions}
The six AIA spectral bands of interest for this study are the 9.4, 13.1, 17.1, 19.3, 21.1 and 33.5 nm bands, all of which are centered on optically thin Fe lines. We focus here on one particular observation, made on 12 March 2011 at 23:55. At that time, Active Region 11166 was producing an uninterrupted sequence of B-class flares and all bands were showing structures with sufficient signal-to-noise ratio. As we shall see later, for BSS to be meaningful, it is important to start with a statistically representative sample of observations that include both active regions and quiet Sun.


Figure  \ref{Figraw}  illustrates the strong similarity between solar images made in the six bands. This similarity is further attested  by Spearman's rank correlation coefficient between each pair of images, which ranges between 0.85 and 0.98. We prefer this correlation measure to the more familiar Pearson correlation because it is invariant to nonlinear rescalings of the intensity. 


In the following, we resample the six calibrated 4k $\times$ 4k images to 2k $\times$ 2k to save computing time; our analysis procedure, however, can be  applied to images of any size. No further preprocessing is done except for dividing each image by its mean absolute amplitude on the disc in order to give comparable weight to all spectral bands. Other rescaling factors, such as the median amplitude, or the median absolute deviation from the mean, give almost identical results.

To extract the sources by BSS, we make the following assumptions:
\begin{itemize}
\item The intensity measured along a given line-of-sight is a linear combination of all different sources (\textit{i.e.} a geometrical mixture). This means that both the sources [$S_k(x)$] and their mixing coefficients [$V_k(\lambda)$] are positive. The latter condition stems from the fact that for each source, all lines must be emitting (\textit{i.e.} non absorbing). This hypothesis is widely supported by observations, \textit{e.g.} \citep{curdt01}.
\item The combination of emissions is linear. This is indeed a reasonable assumption for optically thin lines. However, when appropriate non-linear mixing models are analytically derived, special-purpose BSS methods can be developed \citep{duarte09}.
\item The combination is instantaneous and thus non-convolutive. Any delay in the propagation can indeed be neglected with respect to the detector integration time.
\item Plasma motions are neglected during the interval of observation, which is not the same for all spectral bands.  We checked this by comparing with images taken five minutes before and five minutes after the observation.
\item The observations can be described by a small number of sources, which is akin to saying that the observations are partly redundant. Although this assumption is not mandatory, it enables us to project the observations on a lower-dimensional subspace which, as we shall see, eases their visualization. Many studies support the idea that the Sun-integrated spectral variability in the EUV can be described by a small number of elementary contributions  \citep{lean82,woods00,ddw05,amblard08}. This number is most likely larger for solar images because of the spatial information. Note however that elementary contributions have been reported as well with a set of four SOHO/EIT images \citep{ddw07} and with the modelling of solar structures \citep{feldman10}.
\end{itemize}


\section{Blind Source Separation}

The concept of BSS emerged two decades ago in several disciplines \citep{comon91,jiang04,bobin08} but applications have really taken off only with the advent of robust and fast numerical schemes \citep{comon10}. In space science, BSS has been considered for the exploratory analysis of multispectral astrophysical images  \citep{nuzillard00} but most applications are devoted to the extraction of the cosmic microwave background from \textit{Planck} multispectral images \citep{delabrouille03,leach08}. To the best of our knowledge, the first applications to multispectral solar images were reported by Dudok de Wit and Auch\`ere (2007).

In BSS, neither the sources [$S(x)$] nor the mixing coefficients [$V(\lambda)$] are known \textit{a priori}. Equation~(\ref{bss1}) is therefore heavily underdetermined and the solutions need to be constrained in order to be unique. We assume that the sources present some measurable diversity, which can then be used to disentangle them. In Independent Component Analysis \citep{hyvarinen01}, for example, the sources are forced to be mutually independent. That is
\begin{equation}
\mathcal{P}(S_k,S_l) = \mathcal{P}(S_k) \; \mathcal{P}(S_l) ,
\end{equation}
where $\mathcal{P}(\cdot)$ stands for the probability density function. Independent Component Analysis provides a unique solution with sources that are almost entirely positive and have a clear physical interpretation. \citet{moussaoui08}, however, have shown that the mere independence criterion is not always enough for guaranteeing a proper separation of the sources. Physical reasons in addition lead us to consider only models that enforce $S(x)\ge 0$ and $V(\lambda)\ge 0$.  

From a numerical point of view, the enforcement of the positivity of the sources and the mixing coefficients is major challenge, for which several schemes have recently been developed. Here, we consider one particular approach, called Bayesian Positive Source Separation (BPSS), which has proven to be remarkably efficient and also has the advantage of being deeply rooted in the physics by Bayes' theorem. The algorithm and mathematical aspects such as unicity of the solutions are detailed by \citet{moussaoui06}; an application to Sun-integrated EUV spectra has been made by \citet{amblard08}. 

The solver of BPSS is based on Bayesian estimation theory \citep{gelman11}: we assume that $\mathsf{V} = \left\{ V_k(\lambda) \right\}$ and $ \mathsf{S} = \left\{ S_k(x) \right\}$ are random matrices, whose assessment is to be understood in a probabilistic sense. That is, the problem is solved if we know the posterior distribution
\begin{equation}
\mathcal{P}\left( \mathsf{S},\mathsf{V}\big| \mathsf{I}  \right) = 
\frac{\mathcal{P}\left(  \mathsf{I} \big|  \mathsf{S},\mathsf{V}\right) 
\; \mathcal{P}\left( \mathsf{S},\mathsf{V}\right)}{\mathcal{P}(\mathsf{I})} \ ,
\end{equation}
which is the joint probability distribution of $\mathsf{V}$ and $ \mathsf{S}$, given the data $\mathsf{I}$. This posterior distribution has to be chosen carefully, according to the prior knowledge on the mixing coefficients [$V_k(\lambda)$] and on the source images [$S_k(x)$]. Here we assume that these images and their mixing coefficients are statistically independent random matrices whose distribution is zero for all negative values of their arguments. Typically, we assume that the entries of the matrices are independent random variables, and identically distributed according to Gamma probability density functions. Gamma distributions are frequently used in Bayesian inference as a convenient prior to many likelihood distributions such as Poisson, Gaussian, exponential, \textit{etc.} The elements of the noise term $\mathsf{B}$ are assumed to be independent, zero-mean Gaussian random variables.

We finally obtain the sources and the mixing coefficients from a minimum mean-square-error estimator. The sources, for example, are estimated  as
\begin{equation}
\hat{\mathsf{S}} = \int \mathsf{S} \; \mathcal{P}\!\left( \mathsf{S} | \mathsf{I}\right) \, \textrm{d}\mathsf{S} \ .
\end{equation}
Theses sources and the mixing coefficients can be normalised in different ways. Here, we normalise the mixing coefficients by letting them add up to one for each spectral band.

With six 2k $\times$ 2k images, we find the solution by minimising the error $\mathsf{B}$ in a parameter space with over $2 \times 10^7$ dimensions, using a Markov Chain Monte Carlo method. Such a high dimensionality is a major cost driver since the processing typically requires a hundred iterations, representing about one hour of computer time. As we shall see later, however, there are various ways to turn this powerful technique into an operational tool that can be used in near real-time.

In practice, we fold the 6 $\times$ 2k $\times$ 2k data cube into a (2k)$^2$ $\times$ 6 matrix by lexicographically ordering each image. The relative ordering of the pixels is unimportant as the BPSS does not exploit the spatial structure of the images to refine its solutions. Recent techniques such as morphological component analysis \citep{bobin08,eches11} use the morphological structure as well to improve the image unmixing. Our first tests with such techniques were not conclusive, mainly because solar structures at a given temperature cannot be attributed to a unique morphological class. The hot upper corona, for example, may appear as a diffuse haze (outside of the disc) or on the contrary as thin loops above an active region.


\section{The Analysis Procedure}

Our motivation for using BSS with AIA images is to extract a small subset of source images that are less redundant than the original ones and in this sense concentrate the salient morphological features of the solar corona. As mentioned before, this is justified by the broadband temperature response of the six spectral bands, each of which has a significant overlap with the other bands, see Figure~\ref{Figresponse}. The redundancy of the original images is best quantified by computing their truncated Singular Value Decomposition (SVD), which is one of the simplest and oldest BSS techniques. The truncated SVD consists in applying Equation~\ref{bss1} while constraining the functions to be orthonormal 
\begin{equation}
I(\lambda,x) = \sum_{k=1}^{N_s} A_k V_k(\lambda) S_k(x) + B(\lambda,x) ,
\end{equation}
with
\begin{equation}
\langle V_k(\lambda) V_l (\lambda) \rangle_{\lambda} = \langle S_k(x) S_l(x) \rangle_x = \left\{
\begin{array}{ll}
1 & \textrm{if } k=l \\
0 & \textrm{else}
\end{array}
\right. ,
\end{equation}
where $\langle \ldots \rangle_z$ denotes averaging over variable $z$, and $A_k \ge 0$ are weights.

\begin{figure}[!htb]
\centerline{\includegraphics[width=0.6\textwidth,clip=]{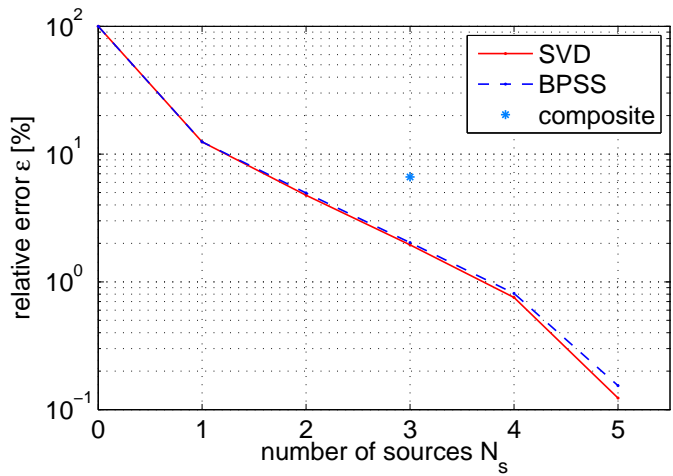}}
\caption{Residual error, as obtained from the SVD and by BPSS. Composite refers to the classical composite image, in which the 17.1, 19.3, and 21.1 nm bands are used as sources.}
\label{Figerr}
\end{figure}

The truncated SVD provides the most compact decomposition (in a least-squares sense) of the original images $I(\lambda,x)$ into a reduced set of $N_s$ separable functions $S_k$ and $V_k$ \citep{cline06}. It thus constitutes a benchmark for testing how well other techniques succeed in decomposing the data into a subset of sources. For $N_s=6$, this decomposition corresponds to the (full) SVD of  $I(\lambda,x)$. The SVD has numerous interesting properties (including data compression and noise reduction) but we shall not use it here for its sources and mixing coefficients are not positive. Let us therefore focus on the residual error [$\epsilon$], averaged over all six images, one makes by reconstructing the images with $N_s \le 6$ sources
\begin{equation}
\epsilon(N_s) = \frac{\langle e^2 \rangle_{x,\lambda}}{\langle I^2 \rangle_{x,\lambda}} \qquad
\textrm{with} \qquad
e(N_s) = I - \sum_{k=1}^{N_s} A_k V_k S_k  = B .
\label{eqerror}
\end{equation}
The residual error tells us how much of the variance (\textit{i.e.} the mean squared intensity) is not described by the sources and is plotted in Figure~\ref{Figerr}. With $N_s=0$, the upper bound is by definition 100\%. With one source only, the residual error drops to 12.4\%, which means that over 85\% of the variance of the images can already be described with one single source. With two and three sources, the error drops respectively to 4.7\% and to 1.9\%. We conclude that a major fraction of the information that is contained in the original images can be efficiently represented by a smaller subset of sources. This is a major incentive for performing blind source separation on AIA images.

A question immediately arises: how many sources are there? The question should rather be: how good should the reconstruction of the original data be? There is no single and robust criterion for answering that question, and the residual error can be made arbitrarily small by selecting enough sources. Figure~\ref{Figerr} suggests that three to four sources is a reasonable choice. We shall start by considering $N_s=1$ up to $N_s=5$ sources. Adding a sixth source clearly will not add much information and, in addition, the solutions tend to become unstable with six sources. It should be mentioned that most BSS techniques cannot extract more sources than there are observations (or in our case, images). Only some recent techniques can do so by adding stronger constraints such as sparsity, but this is still a topic of ongoing research.

\begin{figure}[!h]
\centerline{\includegraphics[width=1.0\textwidth,clip=]{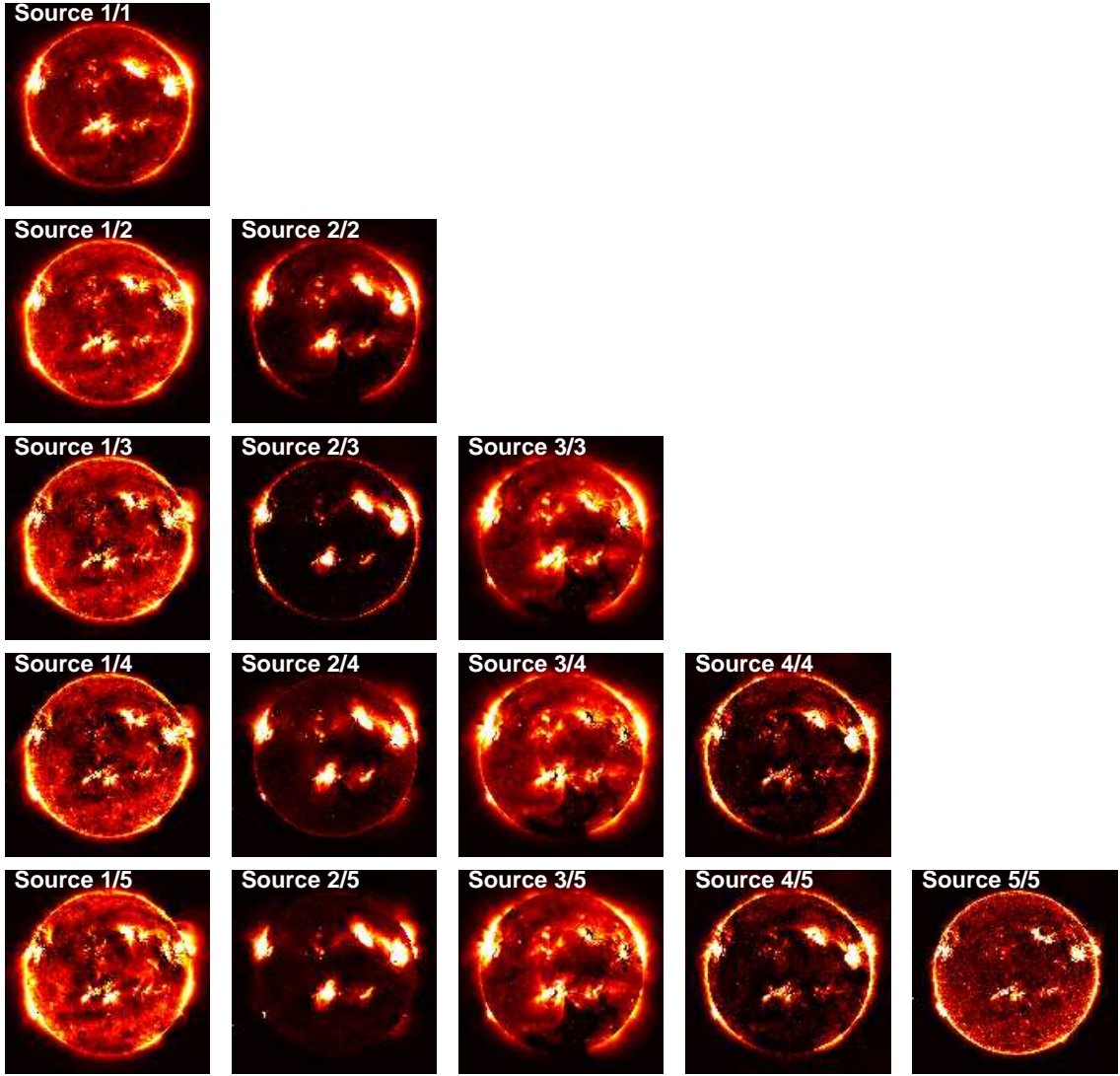}}
\caption{Source images extracted from Figure~1 using respectively $N_s=1$ (upper row) up to $N_s=5$ (lower row) sources. Intensities are shown between the 0.01 and the 0.99 quantiles. }
\label{Figall}
\end{figure}

The result of the BPSS decomposition is summarised in Figure~\ref{Figall}, which shows the source images obtained with $N_s=1$ up to $N_s=5$ sources.  The numbering of the sources is unimportant, so we order them to have similar-looking sources on the same column. Figure~\ref{Figall} reveals several interesting results. First, the contrast between source images is considerably stronger than between the original AIA images; each source now concentrates a specific class of morphological structures that were previously spread out over different spectral bands. This is a natural consequence of the hypotheses behind BPSS; the interpretation will follow in the next Section. 

A second interesting result is the robustness of the sources. The case with $N_s=1$ source is of little interest as the BPSS merely provides an average of all observations. With $N_s=2$ sources, we obtain one source that looks very similar to the 17.1 nm band whereas the other one reveals coronal structures that are best observed in the 9.4 nm band and thus most likely describe the hot corona. These two sources are systematically observed (with some small variants) when $N_s>2$. The same holds for source three out of 3, which also reappears for $N_s>3$, and so on. The third interesting result is that the residual error closely matches that obtained by truncated SVD, which constitutes a lower bound (see Figure~\ref{Figerr}). We conclude that for any number of sources, the observations can always be adequately described as a positive combination of positive sources. This is important, as it confirms the validity of the BPSS model. Finally, let us stress that these results are reproducible in the sense that we obtain the same solutions when running the BPSS solver with different initial conditions.

\section{Interpretation of the Sources}
Our main result so far is the finding of a small number of sources that describe different morphological features in the solar corona. To substantiate this, we shall from now on focus on the case with $N_s=3$ sources, and investigate these more in detail. Let us first consider Spearman's rank correlation between each source and the original images, and between the sources, see Table~\ref{spearman2}. We find that the correlation between sources only is considerably lower than the correlation between original images and sources, which confirms the more pronounced individuality of the latter. Table~\ref{spearman2} also indicates which spectral bands the source are are most correlated with. 

\begin{table}[!htb]
\caption{Spearman rank correlation between original images and the three sources}             
\label{spearman2}      
\centering          
\begin{tabular}{|c|ccc|} \hline
$\lambda$ [nm] & source 1/3 & source 2/3 & source 3/3 \\ \hline
9.4  & 0.804 & 0.662  &  0.857\\
13.1 & 0.923 & 0.514  &  0.767\\
17.1 & 0.968 & 0.427  &  0.717\\
19.3 & 0.792 & 0.509  &  0.948\\
21.1 & 0.697 & 0.545  &  0.971\\
33.5 & 0.749 & 0.590  &  0.932\\ \hline
source 1/3 & 1 & 0.383 & 0.613 \\
source 2/3 & 0.383 & 1 & 0.485 \\
source 3/3 & 0.613 & 0.485 & 1 \\ \hline

\end{tabular}
\end{table}

\begin{figure}[!htb]
\centerline{\includegraphics[width=1.0\textwidth,clip=]{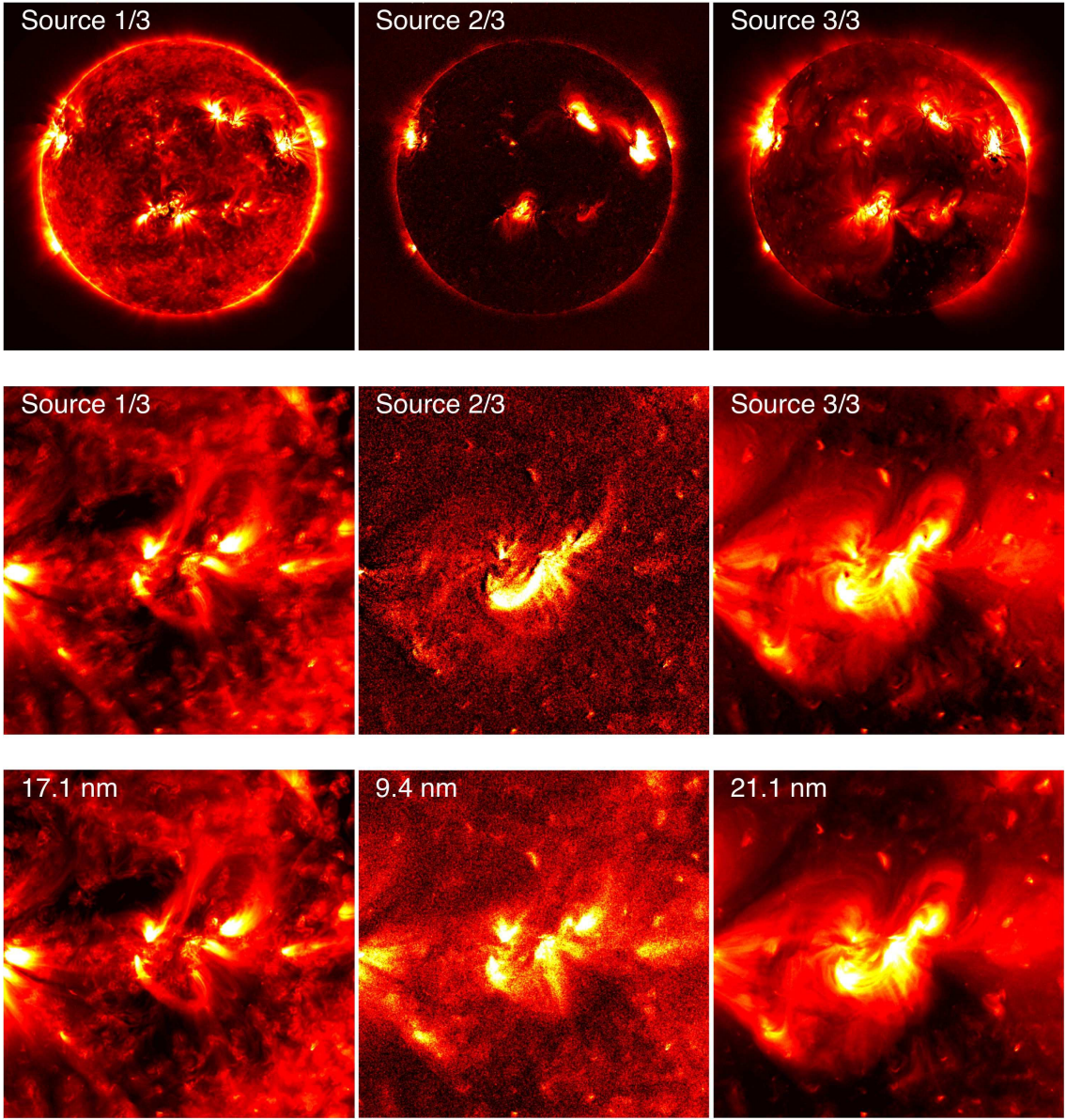}}
\caption{Source images extracted from Figure~1, using three sources. The middle row shows an excerpt and the bottom row compares original images from the spectral band that is most strongly correlated with each source. Intensities are shown between the 0.01 and the 0.99 quantiles.}
\label{Figcompare}
\end{figure}

We find that:
\begin{itemize}
\item source 1/3 is highly correlated with the 17.1 nm band and thus mostly describes the lower corona and upper transition region;
\item source 2/3 is mostly correlated with the 9.4 and the 33.5 nm bands. However, the level of correlation is considerably lower than for sources one and three; this source mostly describes structures that occur at hot active and flaring regions;
\item source 3/3 is mostly correlated with the 19.3 and the 21.1 nm bands and essentially describes the corona and active regions.
\end{itemize}
These results are illustrated in Figure~\ref{Figcompare}, which shows the three sources, an excerpt of a small active region, and the spectral band they are most strongly correlated with. Clearly, source one is similar in all aspects to the 17.1 nm band except for some bright active regions, in which the contribution from the hottest corona is less apparent; it is also more contrasted than the 17.1 nm band. Source three also looks very similar to original images except that it is less contaminated by emissions coming from the lower corona. In particular, the large coronal hole in the southern hemisphere appears more distinctly. 

Source two is the most interesting one because it captures bright structures from the hot corona that are barely apparent in the original images. Note for example how the bright sigmoidal loop stands out. This loop is present in several bands, but it is mostly hidden by other, and most likely cooler, structures. So, not only does BPSS extract sources that are more contrasted than the original images, but in addition it unravels  structures that may go unnoticed. The temperature response of these sources appears to be narrower than that of the original images. Unfortunately, neither \textit{Hinode}/XRT nor GOES/SXI were operating on that day, so that the particular signature from source three cannot be compared with its counterpart in the soft X-ray band. 

\begin{figure}[!htb]
\centerline{\includegraphics[width=1.0\textwidth,clip=]{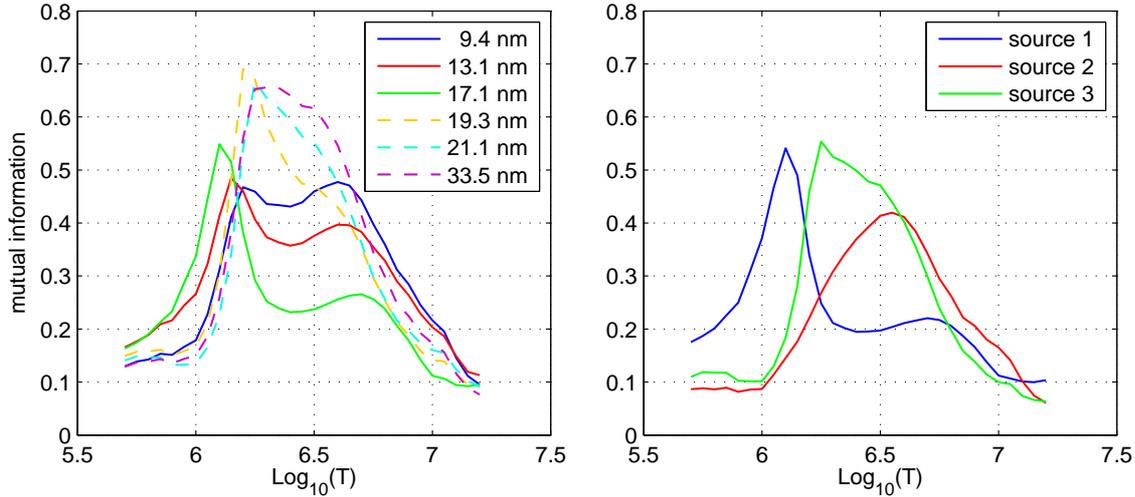}}
\caption{Mutual information between the emission measure at different temperatures and the six original images (left), and the three source images (right). The mutual information is estimated using histograms with eleven equiprobable bins. Values of the mutual information below approximately 0.1 should be discarded because of finite sample effects.}
\label{Figmutinfo}
\end{figure}

To put these results on firmer ground, we estimate the Differential Emission Measure (DEM) from this particular event and compare it to the source images. For each pixel, a DEM distribution that is Gaussian in temperature space \citep{guennou12} is combined with the instrument temperature response functions shown in Figure~\ref{Figresponse} to reproduce the observations. A chi-square test shows that the Gaussian shape is a reasonable assumption except for the few hottest active regions and just above the limb, where multithermal distributions are more likely to occur. The temperature of the best-fit Gaussian DEM is kept for later use. We first reconstruct from the model the emission measure at different characteristic temperatures and compare it to the image intensities. Since the relation between the two is nonlinear, we quantify their correlation by using the mutual information rather than more classical correlation coefficients. The mutual information measures the amount of information that can be inferred about one image, by observing the other, and is defined as \citep{kraskov04}
\begin{equation}
I(X,Y) = H(X) + H(Y) - H(X,Y) \ ,
\end{equation}
where
\begin{equation}
H(Z) = - \int \log z \; \mathcal{P}(z)\; \textrm{d}z
\end{equation}
is the Shannon entropy. The mutual information between two images is zero if and only if these images are independent in a probabilistic sense; it is positive otherwise. Here, we estimate the mutual information by binning image intensities into eleven equiprobable bins. This number provides a good compromise between bias and variance, and the maximum possible value for $I_{X,Y}$ (when $X$ and $Y$ are fully dependent) is then $\log(11) = 2.39$. What really matters, however, is the temperature dependence of the mutual information between each image and the emission measure, see Figure~\ref{Figmutinfo}. This figure and the temperature response shown in Figure~\ref{Figresponse} convey the same message, although they are obtained by completely different means. For example, both show the low temperature peak of the line at 17.1 nm, the bimodal response of several bands, etc. The strong overlap between the curves attests once again the redundancy of the original images. The important results are in the right plot, which shows that: \begin{itemize}
\item the source images have a narrower temperature response than the original ones, and
\item each source image peaks at a different temperature, which has relatively little overlap with the other sources. 
\end{itemize}
From this we conclude that source images are more likely to capture specific morphological structures of the corona because they isolate a relatively narrow temperature band: source one peaks around $\log_{10} T = $6.1 MK, source two around 6.5 MK and source three around 6.3 MK.  Similar conclusions were reached by Dudok de Wit and Auch\`ere (2007), but with three coronal channels of EIT only, which resulted in a much coarser temperature response.

If we now consider $N_s=4$ sources instead, then sources one to three remain almost unchanged, whereas the new source four is mostly correlated with the 9.4 nm band. Figure~\ref{Figall} shows that sources two and four mainly differ by a stronger limb brightening in the latter. We interpret this as a nonlinear effect in the sense that the relative contributions from cold and hot emissions is not only band but also position dependent, in particular when deeper plasma columns are probed just above the limb. Source four tries to compensate for this effect. 

\begin{figure}[!htb]
\centerline{\includegraphics[width=0.65\textwidth,clip=]{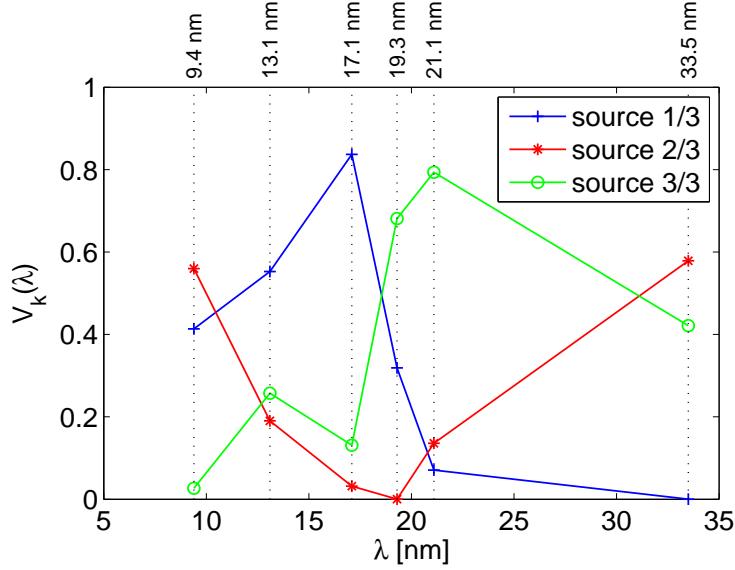}}
\caption{Mixing coefficients obtained with three sources. For each spectral band, the sum of the coefficients is normalised to one. }
\label{Figmix}
\end{figure}

A different aspect of the sources is revealed by looking at the mixing coefficients that describe the fractional abundance of each of them in the observations.  Figure~\ref{Figmix} confirms that the 17.1 nm band mostly consists of source one whereas the 21.1 nm band is dominated by source three. As expected, the 17.1 nm band is devoid of source two. Surprisingly, source two does not appear either in the 19.3 nm band, which should receive a significant contribution from the hot Fe {\sc xxiv} line. The active regions we observe are most probably too cold to be seen in such spectral lines. This highlights the importance of using a representative set of events to estimate the mixing coefficients. During periods of low solar activity, for example, the 9.4 and 13.1 nm bands are known to dominated by emissions that are predominantly coming from the lower corona. If the BSS were to be trained with such data only, the interpretation of these coronal lines would be different.

\section{Temperature Maps}

\begin{figure}[!htb]
\centerline{\includegraphics[width=1.0\textwidth,clip=]{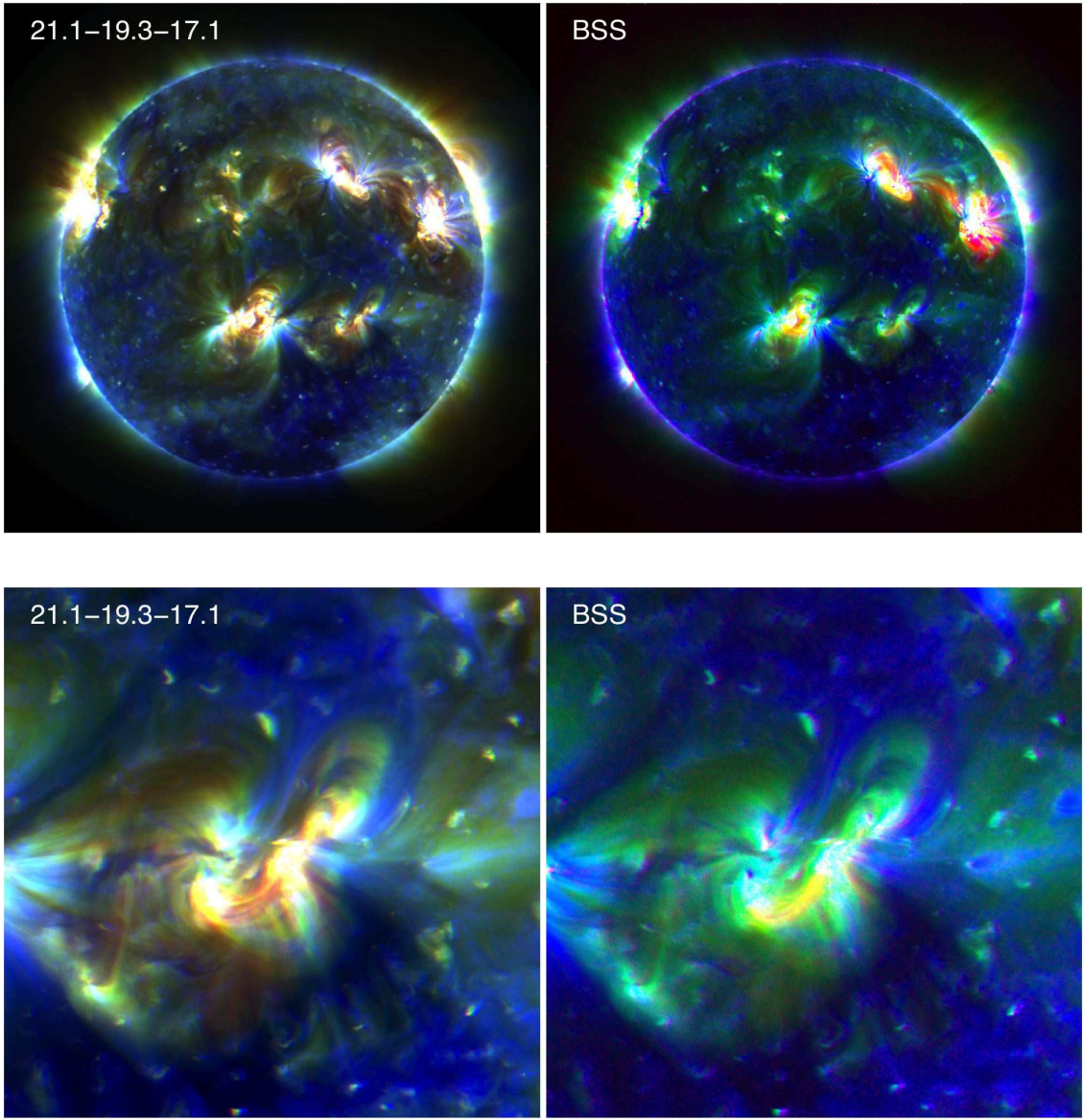}}
\caption{Composite image of the solar corona, obtained by assigning to the red/green/blue channels images at 21.1 nm, 19.3 nm and 17.1 nm (left), and sources two, three and one (right plot). The lower row shows the same excerpt as in Figure~\ref{Figcompare}.}
\label{Figcomposite}
\end{figure}

Now that a temperature range can be assigned to each source image, we  build qualitative temperature maps of the solar corona. This can be done in various ways. The thermal structure is often visualised by making composite images in which three spectral bands are assigned to the red, green, and blue channels. Figure~\ref{Figcomposite} shows such a composite image, in which the 21.1, 19.3, and 17.1 nm bands are respectively assigned to the red, green, and blue channels. The main shortcoming of such composites is the greyish haze that stems from the high correlation between the different spectral bands. This haze is omnipresent near active regions, where all spectral bands receive significant contributions. The BPSS unmixes them and thus provides more contrasted images while incorporating more information, since the information from all six spectral bands is incorporated in the sources. This is confirmed by the residual error, which is 2.0\% for the case with three BPSS sources and at best 6.6\% when three spectral bands are chosen instead (here 21.1 nm, 19.3 nm, and 17.1 nm).

\begin{figure}[!htb]
\centerline{\includegraphics[width=1.0\textwidth,clip=]{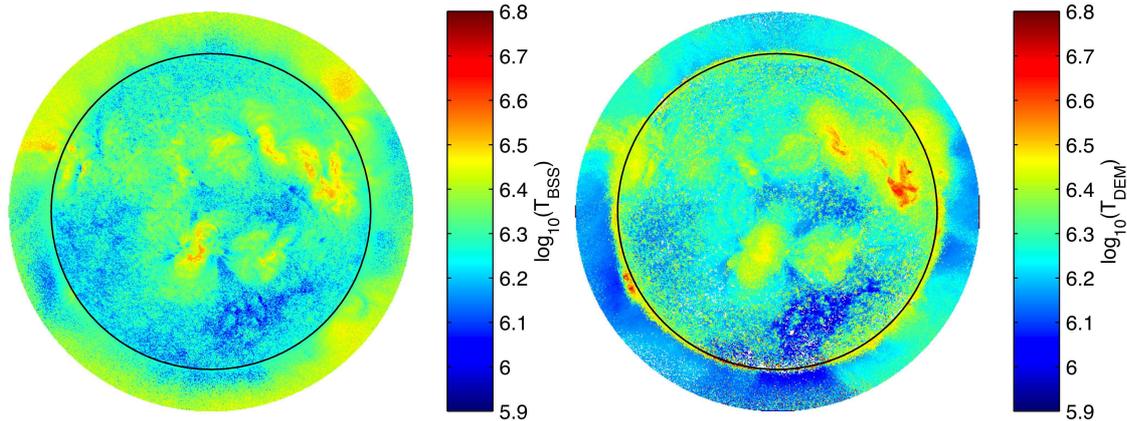}}
\caption{Temperature maps (in $\log_{10} T$[K]) obtained by weighted average of the source images (left) and from the Gaussian DEM model (right). The solar limb is shown by a black circle.}
\label{Figtempmap}
\end{figure}

Figure~\ref{Figcomposite} combines in the same plot information about intensity and temperature. To show temperatures only, we build empirical temperature maps  by assigning to each source its peak temperature: if $S_k(x)$ stands for the intensity of source $k$ at each pixel, then the empirical temperature is defined as the weighted mean 
\begin{equation}
T(x) = \frac{\sum_k T_k \; S_k(x)}{\sum_k S_k(x)} \ .
\end{equation}
Here, $\log_{10} T_1 = 6.1$ MK, $\log_{10} T_2=6.5$ MK and $\log_{10} T_3=6.3$ MK. The resulting map and the one obtained from the peak temperature of the DEM model are compared in Figure~\ref{Figtempmap}. The agreement between both maps is remarkably good, given the fact that they are obtained by completely different means. Their correlation is high on the solar disc (Pearson correlation coefficient = 0.82), but there are also some discrepancies. The DEM model, for example, detects higher temperatures just above the solar limb. There are at least two reasons for this. First, regions above the limb are more likely to exhibit multi-thermal (\textit{i.e.} non-Gaussian) DEMs because of their larger optical thickness. Second, our BSS model does detect a higher temperature above the limb, but that information is contained in source four, which is not considered here. Notice also that the low temperature of the coronal hole near the south pole appears more evidently in the DEM model, because the BSS model cannot correctly represent temperatures that fall outside of the range spanned by $[T_1, T_2, T_3]$.

The key result here is the possibility to rapidly obtain temperature maps from the source images rather than from physical models that are computationally more demanding and require careful calibration.

\section{Outlook}

Although our temperature maps are qualitative and should not be used for calibration purposes, they have several immediate applications. First, they provide convenient quicklook representations because each map condenses in one single picture the salient information that is contained in six AIA images. The temperature maps should therefore be regarded as a powerful data reduction method that reveals the thermal and morphological structure of the solar corona with better contrast. The BSS model is linear, and so one may actually recover from it quantities such as the DEM of the sources. A second application is the segmentation of solar images into regions that provide different contributions to the solar spectral variability. We are presently considering these maps as a direct input to the reconstruction of the solar spectrum in the EUV. 

The analysis could in principle be extended to the chromospheric 30.4 nm, 160 nm, and 170 nm bands, and to the photospheric 450 nm band from AIA. These bands, however, capture morphologically completely different structures such as plages, faculae, umbrae and the network. For that reason, there is no point in adding them in the analysis as they will require additional sources anyway.

Future developments of this promising BSS concept are now being undertaken along three different directions. First, the results need to be put on a firmer basis. This will be done by computing the DEM of the source images by using a non-parametric Bayesian iterative method \citep{goryaev10}. Second, careful validation is needed in order to check the robustness of the results. The BPSS, like the SVD, is data adaptive. However, if the method is applied to different sets of images that are preprocessed in the same way and with the same renormalisation, and if these images contain a blend of structures of different temperatures (including flares), then the mixing coefficients become time-independent. We checked this by analysing different images together and also by computing the BSS for different quadrants of the Sun. The relative standard deviation of each mixing coefficient, as derived from the BPSS model, rarely exceeds 1\%. The relative error, as determined from different images or quadrants is usually 2 to 5 times larger. Whether this time-independence holds on solar-cycle time scales is still an open question. A more systematic study is now underway to quantify this constancy on time scales of months. The only problem that has no workaround is the occurrence of saturated pixels during flares;   source reconstructions then fail locally.

The third issue is the transition to a more operational tool for automated and near real-time use. The main drawback of Bayesian BSS techniques is the computational burden of their Markov Chain Monte-Carlo solver, which presently excludes a systematic analysis of long sequences. The speed of convergence, however, can be accelerated in several ways: First, since changes from one image to another are small if not minute, the latest solution can be used as an initial condition for the next run. Second, as the mixing coefficients are constant in time, the BSS problem actually becomes a problem of solving a linear set of equations with positivity constraints, which is considerably easier to solve. Different solutions have been developed for that purpose, which are presently under investigation. The automation of the process is not an issue as the only important tuneable parameter is the number of sources.

To conclude, the transition toward an operational tool is a challenging task but is within reach. The optimisation of Bayesian BSS today is a major issue, but that field is rapidly evolving \citep{kuruoglu10}. In the present study we wanted to focus instead on the concept and its physical interpretation.

\section{Conclusions}

This study shows that the BSS is a novel and powerful concept for easing the analysis of solar EUV images within a Bayesian framework. The key results are: \textit{i.}) BSS allows us to condense in one single picture the information that is contained in multiple spectral bands, \textit{ii.}) the sources this picture is made of have an immediate physical interpretation as they describe specific temperature bands of the solar corona and capture morphologically different structures, and \textit{iii.}) together, these sources provide a more contrasted picture of thermal structuring of the solar corona than the original images, which always contain a blend of emissions coming from different temperatures. 

The source images we obtain by BSS are empirical as they are derived from the statistical properties of the images only. However, since they are obtained just by linear combination of the original images, they can serve as inputs to semi-empirical models. The computational complexity of Bayesian methods is a real challenge, but not a major obstacle. We believe that the enormous potential of Bayesian BSS makes of it a powerful concept for analysing multi-wavelength solar images.

%
\subsection*{Acknowledgements}
TD and MK gratefully acknowledge the International Space Science Institute (ISSI, Bern) for hospitality.  This study received funding from the European Community's Seventh Framework Programme (FP7/2007-2013) under the grant agreement number 218816 (SOTERIA project, \url{www.soteria-space.eu}) and from the Programme National Soleil-Terre (PNST). We also thank Thomas Benseghir and Nolwenn Marchand for their assistance in the data analysis. The AIA data are courtesy of SDO (NASA) and the AIA consortium.

%
%

\end{document}